# Optical characterization of a-Si:H thin films grown by Hg-Photo-CVD

**A. BARHDADI\*,  S. KARBAL,  N. M'GAFAD**

*Laboratoire de Physique des Semi-conducteurs et de l'Energie Solaire (P.S.E.S.)*
*Ecole Normale Supérieure, P.O. Box: 5118, Takaddoum, Rabat 10000, Morocco*
*Unité de Formation et de Recherche en Physique de la Matière Condensée et Modélisation*
*Statistique des Systèmes, Faculty of Sciences, University Mohammed V-Agdal, Rabat 10000, Morocco*
*\*The Abdus Salam International Centre for Theoretical Physics (ICTP), 34014, Trieste, Italy*

&

**A. BENMAKHLOUF,  M. CHAFIK EL IDRISSI**

*Equipe de Physique des Surfaces et Interfaces (E.P.S.I.)*
*Laboratoire de Génie Physique et Environnement*
*Unité de Formation et de Recherche en Optoélectronique et Spectroscopie*
*Faculty of Sciences, University Ibn Tofail, P.O. Box: 133, Kenitra 14000, Morocco*

&

**B. M. AKA**

*Département des Sciences et Technologie, Ecole Normale Supérieure, University Abobo-Adjamé*
*P.O. Box 1561, Abidjan 22, Côte d'Ivoire*

\* *Corresponding author (Senior Associate of the Abdus Salam ICTP)*
 *Phone: (212) 37 75 12 29  or  (212) 37 75 22 61 or  (212) 64 93 68 15*
 *Fax: (212) 37 75 00 47*
 *E-mails: barhdadi@ictp.it   or   abdelbar@fsr.ac.ma*

# Abstract

Mercury-Sensitized Photo-Assisted Chemical Vapor Deposition (Hg-Photo-CVD) technique opens new possibilities for reducing thin film growth temperature and producing novel semiconductor materials suitable for the future generation of high efficiency thin film solar cells onto low cost flexible plastic substrates. This paper provides some experimental data resulting from the optical characterization of hydrogenated amorphous silicon thin films grown by this deposition technique. Experiments have been performed on both as-deposited layers and thermal annealed ones.

# Résumé

La technique de dépôt photochimique en phase vapeur sensibilisé au mercure (Hg-Photo-CVD) ouvre de nouvelles possibilités pour réduire la température de croissance des couches minces et produire de nouveaux matériaux semiconducteurs convenables pour la future génération de photopiles, de haut rendement et petit prix, réalisées en couches minces déposées sur des substrats flexibles en plastique. Ce papier présente des résultats expérimentaux obtenus à travers la caractérisation optique de couches minces de silicium amorphe hydrogéné élaborées par cette technique de dépôt. Les mesures ont été effectuées en faisant varier soit la température de croissance soit la température de recuit.

# I. INTRODUCTION

The key driver for the next future generation photovoltaic is the cost reduction to about $ 1/W_p$ at the system level. In order to achieve this ambitious goal, the focus of photovoltaic R&D is currently in developing thin film technologies, involving new semiconductor materials and solar cell devices [1, 2]. Thin film technologies hold, indeed, considerable promise for a substantial reduction of the manufacturing price of solar cells due to the reduction of materials cost and their possible deposition on large area substrates [3 - 7].

Thin film technologies for PV applications include various semiconductors materials and conductive oxides [8]. The most advanced of them are those using Hydrogenated amorphous Silicon (a-Si-H), Cadmium Telluride (CdTe) and CIGS materials [3, 4]. Nevertheless, solar cells and a range of other electronic devices technologies based on a-Si-H come in the first rank because of the many specific advantages offered by this interesting semiconductor material. Indeed, a-Si-H is an extremely abundant raw material and involves almost no ecological risk during purification and processing. Also, with comparison to crystalline silicon, a-Si:H is distinguished by a large optical gap in the required range for optimal PV conversion [9]. It presents also a high optical absorption within the maximum of solar spectrum. Therefore, a large part of solar energy is absorbed in a small thickness of the material allowing making structures in the form of very thin layers. So, a thickness of typically 1 μm material suffices to efficiently absorb most of solar radiations. Moreover, deposition of a-Si:H is much faster than crystalline silicon growth and can be carried out over much larger areas as mentioned above. Low process temperatures facilitate the use of a variety of low cost substrate materials such as float glass, metal or plastic foils [10 - 12]. All these characteristics are of great importance making a-Si:H a more attractive semiconductor material for the elaboration of cheap and flexible photocells having good photovoltaic parameters.

The performances of a-Si:H solar cells are intimately related to the film structure determined by the preparation method and the experimental conditions adopted during the deposition process. Various deposition techniques have been employed [13 - 18]. Among them, Plasma-Enhanced Chemical Vapor Deposition (PECVD) [18 - 23] still the most commonly used for the direct growth of a-Si:H thin films on low substrate temperature, and $SiH_4$ heavily diluted with $H_2$ are generally used as the reactants. However, PECVD has yet two serious drawbacks: the surface damages which result from the impinging charged particles with high energy of the plasma, and the impurities incorporation which results from sputtering because of the high potential difference between substrate and electrodes in the reactor [24].

In our previous work [25 - 29] we have prepared a-Si:H thin films by radio frequency cathodic sputtering technique and we have studied how optical characteristics of these films change with increasing hydrogen pressure during the deposition stage as well as with classical post-deposition annealing [25 - 27]. Using the grazing X-rays reflectometry technique, we have also characterized some of the structural properties of a-Si:H very thin layers immediately after deposition as well as after surface oxidation or annealing [28, 29].

Recently, Myong et al. [30] reported that Mercury-Sensitized Photo-Assisted Chemical Vapor Deposition (Hg-Photo-CVD) is an efficient method for deposing high quality a-Si:H thin films at temperature as lower as 120˚C. They showed also how this technique is promising for developing an innovative generation of nano-crystalline silicon (nc-Si) thin film solar cells fabricated onto low cost flexible plastic substrates. At home laboratory, we have been very interested by Myong et al. achievements and we have agreed to develop this new topic among our current research activities to eventually perform some significant original contributions. So, we started working on this topic through the conception and realization of a new a-Si:H thin films deposition set up based on the Mercury-sensitized photo-CVD technique [31 - 33]. This set up, which is progressively developed in home laboratory, is quite similar to that performed by Aka some years ago in France [34 - 37]. The specific research program we are currently carrying out on this new system registers within the framework of a research project financed by the Moroccan ministry of high education and scientific research. It mainly consists in depositing good quality a-Si:H and nc-Si thin films in the perspective to contribute in the development of the future generation of low cost powerful terrestrial solar cells. To conduct well this research program, we initially performed an extensive bibliographical work aiming to seek and join together most of scientific information and technical data published by the specialists of the subject. After examining and analyzing these literature data, we noted some fundamental points characterizing particularly both a-Si:H and nc-Si materials and Hg-Photo-CVD technique. That is what we have tried to recall and review through the bibliographical synthesis we have published in recent papers [38 - 41]. Our motivation was to develop a new synthesis work dedicated mainly to outline the great photovoltaic potential of the low temperature deposition process of a-Si:H [38 - 39] and nc-Si [40 - 41] thin film materials for the development of the new generation of high efficiency low cost solar cells.

In the present work, we provide the readers with some experimental results we have obtained on this subject. We have prepared a-Si:H thin films using Hg-Photo-CVD technique and we have studied how optical characteristics of these films change with increasing the substrate temperature during the deposition and what effects may be observed from a post-deposition thermal annealing.

## 2- EXPERIMENTAL DETAILS

### 2-1- Description of Hg-Photo-CVD system

The Hg-photo-CVD set up we have developed in home laboratory is very simple. Its schematic diagram (figure 1) was described elsewhere [38 - 41]. It consists mainly of a cylindrical horizontal quartz reactor in the form of a special tube with size and dimensions as specified in figure 2. This tube, used as deposition chamber, has two access gates for the entries and exits of gases. It is equipped with a small reservoir containing a small quantity of liquid mercury (Hg bath) which can be thermally controlled independently of the remainder of the system. The inner surface of the tube may be coated with low-vapor-pressure Fomblin vacuum oil to prevent any film deposition on [42 - 44]. The substrate support has a rectangular form and can be of graphite or of stainless steel. It is heated through a thermo-coax wire and equipped with a Chromel-Alumel thermocouple to measure the substrate temperature $T_s$ in a wide range [- 200˚C, + 1000˚C]. Well-cleaned high resistivity crystalline Si wafers or Corning 7059 glass can be used as substrates. The distance between the tube inner surface and substrates is about 3 cm. The UV light source consists of a series of PHILIPS-TUV low pressure Hg lamps radiating both 253.7 nm (40 mW/cm$^2$ or ~ 30 mW/cm$^2$ at 3 cm distance) and 184.9 nm (less than 10 mW/cm$^2$ or ~ 5 mW/cm$^2$ at 3 cm distance) resonance lines. Since the transmittance of the quartz tube for the 253.7 nm and 184.9 nm wavelengths is 82 % and 20 % respectively, the 253.7 nm resonance line of UV light is dominantly irradiated into the reactor. The series of Hg lamps are set up under an aluminum reflector placed at about 5 cm distance from the reactor. The vacuum system is composed of two pumps. The first one, with pallets, is used to obtain a primary vacuum and to purge gases of the reactor. The second is a diffusion pump allowing evacuation down to a pressure of 10$^{-6}$ Torr during the back out of the reactor prior to the growth. The inside total gas pressure is adjusted by a conductance valve and measured by a MKS Baratron capacitance manometer placed rightly in the entry of the reactor. This manometer has the advantage to indicate pressure values independently to the nature of gas used, contrary to the Pirani jauges that we have associated to the vacuum pumps. Reactant gases, mainly SiH$_4$ and H$_2$, are introduced into the reactor through the Hg bath usually kept at temperature between 35°C and 50°C. Thus, a very small amount of Hg vapor is automatically mixed with the gases and introduced into the deposition chamber. The Hg vapor atoms, then introduced, enhance the dissociation of $SiH_4$ [45] because of its weak optical absorption in the 190-200 nm wavelengths region [46, 47].

Because our Hg-photo-CVD set up is quite similar to that performed by Aka [34 - 37], we suppose that experiments we may conduct on both systems in the same conditions should provide the same results.

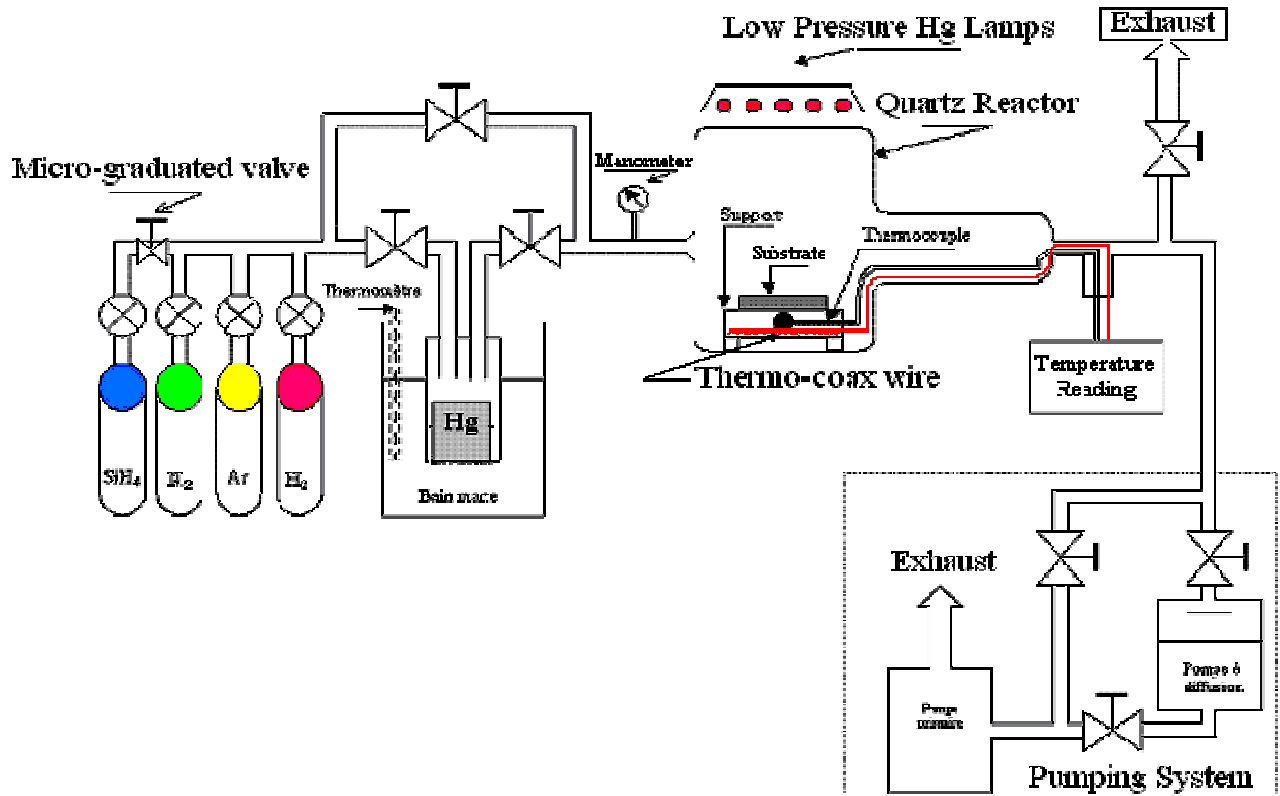

*Figure 1: Schematic diagram of the Hg-sensitised photo-chemical vapor deposition set up developed at home laboratory*

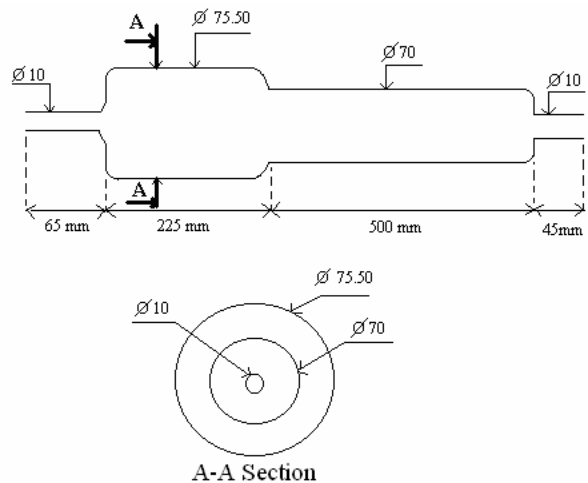

*Figure 2: Schema and size dimensions of the cylindrical horizontal quartz reactor used for the Hg-sensitized photo-chemical vapor deposition set up developed at home laboratory*

## 2-2- Sample preparation

The a-Si:H thin film samples were deposited from the photodissociation of $SiH_4$ reactant gas mixed with $N_2$ buffer inert gas. This dilution is required for both economical and safety reasons since the presence of $N_2$ at high enough concentration eliminate the explosive character of some hazardous reactions as specified in the literature [48]. The gas mixture was introduced into the reactor through the Hg bath to carry a very small amount of mercury vapor. The $SiH_4$ flow is controlled by a micro-valve. The substrates used are of Corning 7059 glass with a small surface of about 1 $cm^2$. Before any deposition, they have been submitted to an appropriate cleaning to prevent eventual contaminations. The substrate temperature ($T_s$) being the main parameter that we intented to vary, substrates were then inserted in the reactor and heated up the desired temperature ranged from 100°C to 400°C by step of 100°C. Table 1 summarizes all our experimental growth conditions. They have been chosen on the basis of results deduced from previous works [34]. After 30 min deposition run, thin films obtained are with a good adherence. They are supposed homogeneous and their thickness is uniform and not exceeding 0.5 μm.

| Parameter | Value |
| --- | --- |
| Substrate | Corning glass C 7059 |
| Base pressure | $10^{-6}$ Torr |
| Total deposition pressure (P) | 5 Torr |
| Growth temperature ($T_s$) | Ranging from 100°C to 400°C with 100°C step |
| Silane flow rate | 1 sccm |
| Total lamp power | 2 mW/$cm^2$ |
| Hg bath temperature | 50°C |

*Table 1: Summary of growth conditions*

## 2-3- Experimental analyses

The whole hydrogen concentration as well as its spatial distribution in the films was determined by ERDA technique [49]. Their optical properties were measured by means of a Beckman UV-5270 spectrophotometer operating with a double beam in a wide spectral range (0.25 μm - 3 μm). Transmission measurements were taken by means of the differential method [50] which consists in determining the transmission of the layer together with its substrate by comparison to another identical virgin substrate used as reference. Under these conditions, the influence of the substrate on the transmitted light is practically negligible [50]. Some samples have been also characterized using Infra-Red Spectroscopy measurements to look at how hydrogen atoms are bonding within the lattice.

## 3 - RESULTS AND DISCUSSIONS

Thin films obtained were characterised immediately after deposition without undergoing any particular processing. Some of them were examined before and after being thermally annealed.

### 3 -1- Measurements performed just after thin films growth

These measurements were performed on a-Si:H thin films deposited on substrates whose temperature ranges from $T_s$ = 100°C to $T_s$ = 400°C by 100°C step. In order to determine the total concentration of hydrogen in these as-deposited films, ERDA measurements have been systematically performed on all of them. The results are summarized in the table 1. We can easily note that thin films are in general highly hydrogenated. This could be explained by the $SiH_4$ low flow we have used for the films growth process [34]. We can also clearly note that hydrogen relative content in the films decreases quite regularly according to increase of their growth temperature.

| $T_s$ (°C) | [H] / [Si] (%) |
|---|---|
| 100 | 42 |
| 200 | 30 |
| 300 | 20 |
| 400 | 13 |

*Table 2: Hydrogen relative concentration in the films according to the growth temperature $T_s$*

Figure 3 shows a typical transmission spectrum (normalized to that from the substrate) obtained for the a-Si:H thin film deposited at $T_s = 200°C$. All the other transmission spectra obtained for the different a-Si:H samples deposited at various $T_s$ present practically the same form. They all exhibit two clearly distinct slopes [51]. The first one, in which interference fringes (oscillations) are seen, is commonly named transparency zone or zone of weak absorption. In the second one, which is well known as zone of high absorption, the signal is strongly reduced.

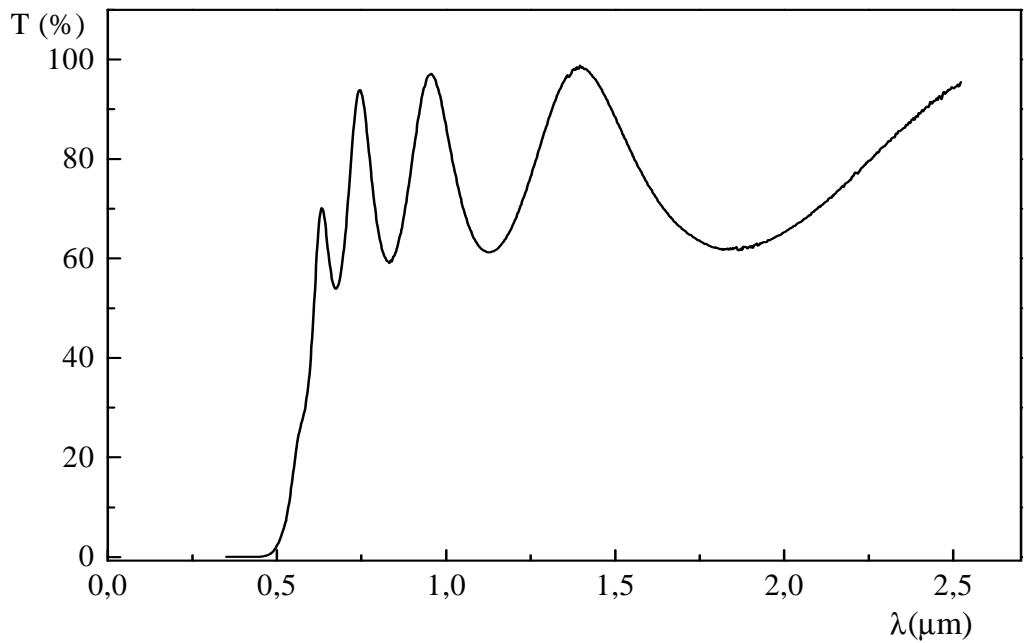

*Figure 3: Transmission spectra, normalised to the substrate, obtained for an a-Si:H thin film deposited at substrate temperature kept at 200°C.*

By exploiting the different transmission spectra corresponding to the various $T_s$ values, and using the mathematics expressions reported in the reference [52], we were able to determine the main optical characteristics of our samples, especially the refraction index (n), the optical gap ($E_g$) and the Urbach energy ($E_u$) [53]. It should be interesting to remember that the latter draws light on the density of the energy states localised in the tail of the valence band, which are generally attributed to the structural disorder in the material [54].

In figure 4, we reproduce the dependence of refraction index n with the wavelength $\lambda$ for all $T_s$ considered. The values of n were adjusted using the dispersion law of Sellmeiere (equation 1) [55].

$$n^2(\lambda) = n_\infty^2 + \frac{b^2}{\lambda^2 - \lambda_0^2} \quad (1)$$

where $n_\infty$ is the refraction index obtained by extrapolation towards the infinite; b and $\lambda_0$ are constants determined by the $n(\lambda)$ curve fitting.

From figure 4, we can note that n values are relatively low but similar to those obtained by both Zarnani et al. [56] and Toyoshima et al [57] in the case of thin films photo-deposited from $Si_2H_6$ and $Si_3H_8$ by means of Excimer laser emitting at 193 nm. It is clear also that, when $T_s$ increases, $n(\lambda)$ curve shifts towards higher n values. This evolution, which has been already observed by many other authors [56, 58 - 60], should be closely associated to the hydrogen concentration in the layers. In others words, when the hydrogen content decreases, the density of dangling bonds in the film increases in parallele with the layer absorption and, then, an enhancement of the refraction index is observed. Indeed, for a given material, the more its absorption is higher, the more light reflection at its surface is better (metallic luster).

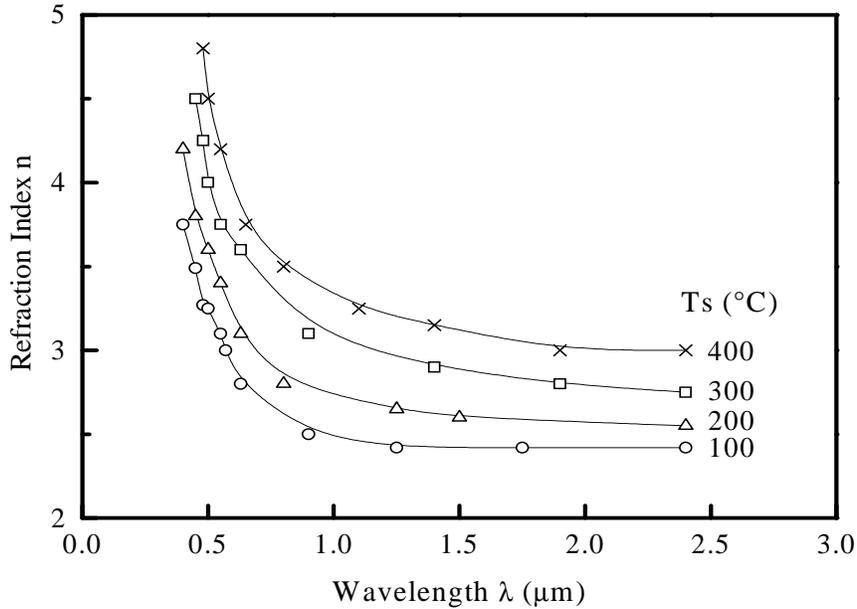

*Figure 4: Experimental curves showing the evolution of the refraction index n of the deposited layers as function of the wavelength $\lambda$ for various $T_s$.*

Figure 5 shows the evolution of $E_g$ and that of $E_u$ as function of $T_s$. $E_g$ values were determined from Tauc formula [61] expressed by equation 2, and those of $E_u$ from Urbach law [53] formulated by equation 3.

$$(\alpha h\nu)^{1/2} = B(h\nu - E_g) \quad (2)$$

$$\alpha(h\nu) = \alpha_0(h\nu) \exp\left(\frac{h\nu - h\nu_0}{E_u}\right) \quad (3)$$

$\alpha$ is the absorption coefficient, $h\nu$ is the photon energy, B is a coefficient of proportionality, $\alpha_0$ and $h\nu_0$ are constants depending on depositing conditions of the films.

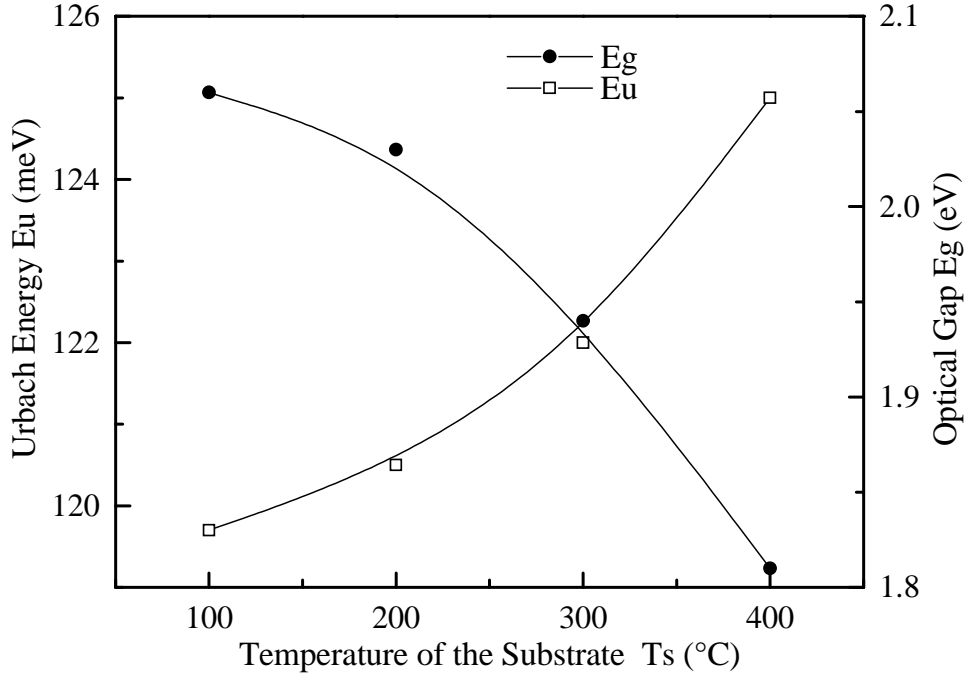

*Figure 5*: Evolution of the optical gap ($E_g$) and of the Urbach energy ($E_u$) measured on deposited layers as function of substrate temperature $T_s$. $E_g$ was determined from the Tauc formula [61] and $E_u$ from Urbach law [53]. We show that both $E_g$ and $E_u$ exhibit strictly opposed evolution trends

We notice that $E_g$ and $E_u$ evolve into strictly opposing directions: when one increases the other decreases. This clearly confirms the very close relationship existing between these two parameters because of the presence of a more or less important density of dangling bonds in the material.

Figure 5 shows also that, when $T_s$ increases, $E_g$ decreases quickly. Co-jointly, $E_u$ perfectly develops in the opposite direction. From these results, we can say that for low $T_s$, hydrogen incorporation saturates dangling bonds in the deposited layers. This leads to a clear improvement of $E_g$ and $E_u$. When $T_s$ = 100°C, the density of dangling bonds is minimal and, hence, the optical performances of the layers are optimal ($E_g$ is maximal and $E_u$ is minimal). When $T_s$ becomes high and higher, the out diffusion of hydrogen from the material should be more and more important. The less presence of hydrogen in the material generates new structural defects. This has negative repercussions on the optical performances of the layers, and leads to a diminution of $E_g$, accompanied by an increase of $E_u$ as seen on figure 5.

### 3 -2- Measurements performed after thermal annealing

After examining the effects of growth temperature on the optical properties of a-Si:H thin films immediately after their photo-depositing, we focused on the changes that would affect these properties after thermal annealing. It is essential to keep in mind that stability of photovoltaic a-Si:H compounds depends on their normal functioning temperature and on the thermal solicitations to which they are sometimes compelled to be submitted. This is why many studies hold on this extremely important aspect of the problem [62 - 65].

The effect of annealing temperature $T_a$ we are studying in this section is especially interesting for removing defects, inherently existing in the amorphous structure, in order to improve the conduction properties of the material.

While heating an a-Si:H sample, at least two main different situations may be observed. In the first one, a reversible annealing effect happens. This means the material optical parameters, such as $\alpha$ and n, recover their initial values at ambient temperature as soon as the heating is stopped. In this case, the material doesn't keep any trace of the annealing effect. In principle, this reversible effect should occur for annealing temperatures lower or closer to the growth temperature $T_s$. In the second possible situation, an irreversible annealing effect happens. This means that heating has modified the material properties and, therefore, measurements at the ambient temperature couldn't give back the initial values of parameters.

In order to conduct this part of work, we have prepared four a-Si:H samples under experimental conditions similar to those mentioned above, using a $T_s$ fixed at 200°C. One of these samples is kept as a control. The other three ones were each submitted to 2 hours isochronal thermal annealing in a conventional furnace under $10^{-6}$ mbar vacuum. Annealing temperature $T_a$ was chosen to be the variable parameter of which we try to determine the effect. $T_a$ was fixed at 340°C for the first annealing, at 400°C for the second and at 570°C for the third one.

Figure 6 shows the evolution of the absorption coefficient $\alpha$ with the wavelength $\lambda$ for each of the four examined samples. The value of $\alpha$ was calculated from the absorption spectra measured on these samples. In this figure, only presented the range of $\lambda$ (0.50 µm ≤ $\lambda$ ≤ 0.65 µm) for which the evolution of $\alpha$ is more interesting. We can easily see that as the annealing temperature increases, the spectrum shifts toward the higher $\lambda$. In other words, $\alpha$ increases clearly with $T_a$ and this increase is more higher when $\lambda$ is more lower. For example, for $\lambda$ = 0.6 µm, $\alpha$ measured after 570°C thermal annealing is 10 times higher that measured on the non annealed layer.

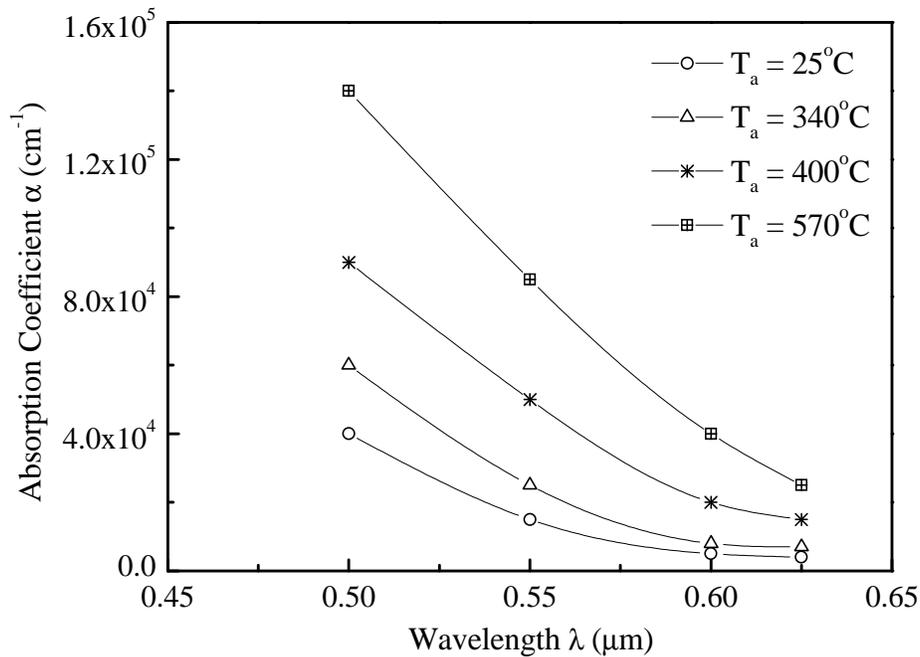

*Figure 6: Evolution of the absorption coefficient $\alpha$ with the annealing temperature $T_a$*

On the basis of experimental results obtained from X-rays analysis, Chahed et al. [63] found that for as deposited a-Si:H thin films as well as for low temperature annealed ones, the film density is distinctly weaker than that of crystalline silicon. Only thermal annealing at temperature as higher as 600°C enhances the densification of the films and reduces significantly their thickness. Therefore, the evolution of thin films optical parameters with thermal annealing should be examined according to the structural modifications of the layers. Basically, these modifications result from some hydrogen out-diffusion during thermal processing allowing the resurgence of structural defects depending strongly on how stable are the various hydrogen configurations in the layers. Indeed, thermal annealing at increasing temperature up to the crystallization stage produces a progressive out-diffusion of hydrogen. However, for the bounded hydrogen atoms, this out diffusion is generally observed for $T_a$ higher than $T_s$ [63, 66]. For the lower $T_a$, thermal processing only induces some homogeneity in the structure of deposited layers without any significant effect on their electronic properties [63].

Figure 7 shows the spectral scattering of the refraction index n for all the annealed samples. We notice that thermal annealing induces a clear n increasing in the whole-explored spectral range. This increase is obviously slight for $T_a = 340°C$ but, when $T_a$ reaches 570°C, it becomes so important that, for high λ, n reaches typical values usually measured on silicon crystals [67]. With regard to this result, we can already say that thermal annealing at 570°C allows a partial crystallisation of the deposited layer. Hence, we meet literature data which show that amorphous silicon crystallises at a critical temperature between 550°C and 700°C during annealing process [64].

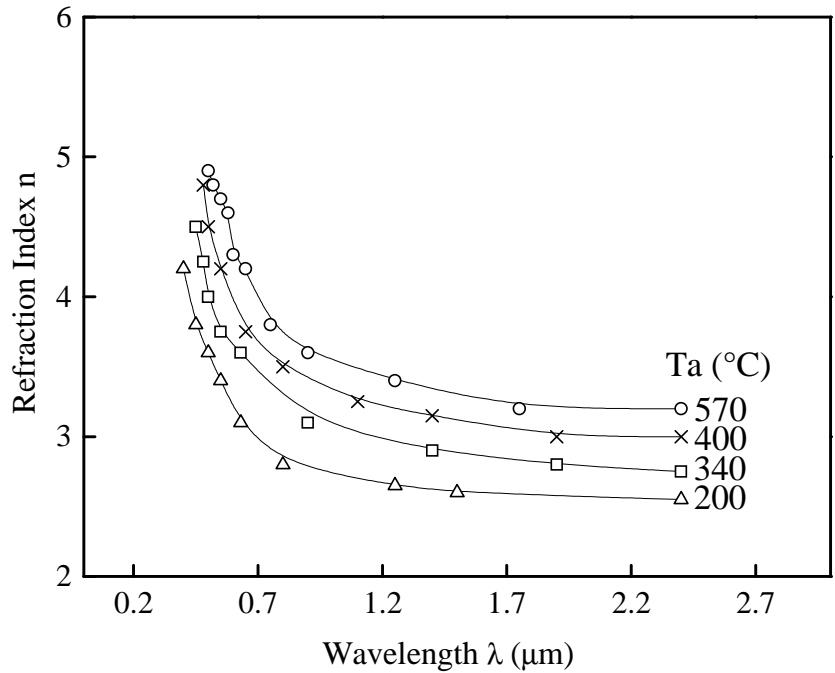

*Figure 7: Spectral scattering of the refraction index n and its evolution with the annealing temperature $T_a$*

At least, two processes can be identified from the variation of the optical gap $E_g$ with the annealing temperature $T_a$. The first one occurs when the annealing temperature is close to the depositing one ($T_a \sim T_s$). It consists in the liberation of hydrogen atoms that are weakly bounded in the matrix, followed by a reconstruction of the material leading for some structural disorder. The second process is the matrix annealing which is generally accompanied with the generation of a high density of dangling bonds because of the broken strong Si-H bonds and the new defects induced in the material by heating and thermal constraints [66, 68, 69].

In figure 8, the variations of $E_g$ and those of $E_u$ with the annealing temperature $T_a$ were both reported on the same graph. The first remark is the close relationship that links these two parameters making their evolution curves quite symmetric. Furthermore, these curves present an important similarity with those found in the literature [34]. Examination of these curves allows noting the following points:

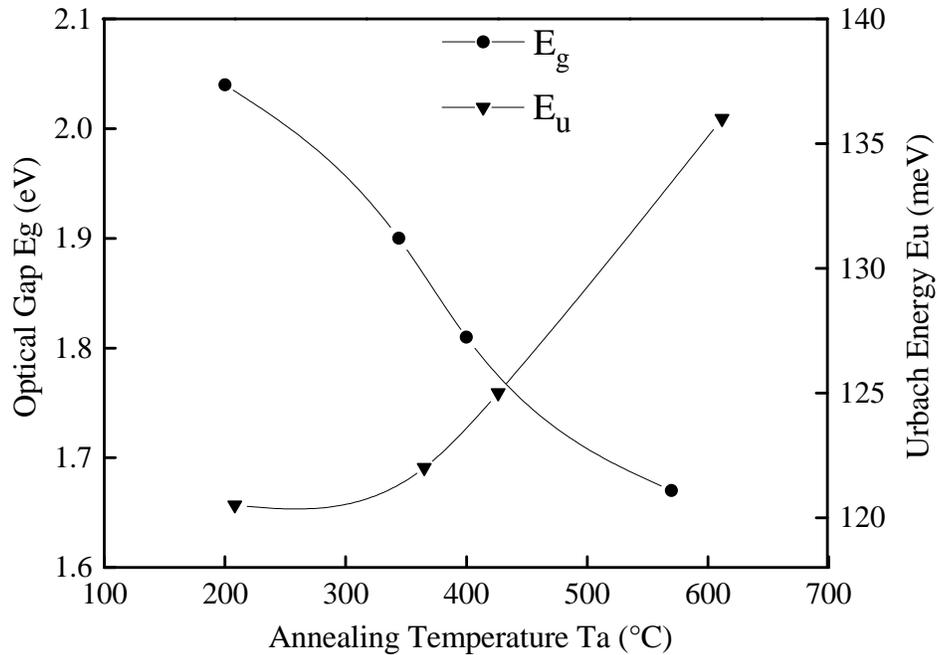

Figure 8: *Variations of the optical gap $E_g$ and the Urbach energy $E_u$ according to the annealing temperature $T_a$. The strong co-relation linking these two parameters makes their curves completely symmetrical*

a) For the non annealed sample (control), $E_g$ and $E_u$ present values practically equal to those measured above on the sample that was prepared in the same experimental conditions (figure 5). This proves that our results are reproducible. In other words, the structural quality of the deposited layers is practically influenced only by deposition experimental conditions which are entirely under control.

b) The irreversible effect starts at 200°C which is in the same time the value of annealing temperature $T_a$ and growth temperature $T_s$ ($T_a = T_s$). Between this temperature and that of thermal annealing at 570°C, $E_g$ decreases from 2.04 eV (which is also the value obtained from the un-annealed sample) to 1.67 eV.

c) Thermal annealing at 340°C or at 400°C simultaneously leads to a decrease in $E_g$ and to a proportional increase in $E_u$. On the basic of literature data [34], we attribute this to a partial out-diffusion of hydrogen resulting from a break of the less stable bonds of $SiH_2$ and $SiH_3$ that have been formed in the material during the growth.

d) After annealing at 570°C, $E_g$ decreases considerably reaching a value of 1.67 eV. Simultaneously, $E_u$ increases to reach an average value of 136 meV. Logically, and in the same way as in the two first annealings operated at 340°C and 400°C, these variations can also be attributed to the out diffusion of a higher quantity of hydrogen as a result of the break in the more stable bonds of Si-H [34, 70]. Nevertheless, this interpretation is not sufficient to well explain the changes we observed, even if we assume that all the quantity of hydrogen incorporated in the material during depositing has out diffused. Indeed, after an annealing at 570°C, $E_g$ and $E_u$ values are very different from those measured on an unannealed sample deposited in the same experimental conditions at a temperature $T_s$ = 570°C. This proves that, even if the out diffusion of hydrogen is total, it cannot explain only by itself the observed results. Taking in consideration the values of $E_g$ and $E_u$ we found, the most plausible interpretation of our results should combine the hydrogen out diffusion phenomenon to an efficient process of crystallisation of the layers, which is well favoured by the high values of $T_a$ [62, 65, 71, 72]. However, the creation of new structural defects in the material because of thermal constraints and heating is not excluded [73].

## 4- CONCLUSION

The present paper is exclusively dedicated to the presentation of the new Hg-Photo-CVD set up we have developed at home laboratory, and the exposition of some experimental results obtained by this kind of thin films photo-deposition systems. We have studied the effect of the substrate temperature $T_s$ during the growth process, as well as the post-growth thermal annealing temperature $T_a$, on the optical parameters of a-Si:H thin films photo-deposited by this technique. Usually, in the photo-assisted process, the depositing rate increases with the intensity of UV light generated by the source of photons. In our study, because we have used the available low pressure Hg lamps as UV source, thin films growth rate was relatively low. The obtained results show that as deposited thin films hold a very high quantity of hydrogen (about 40% for $T_s$ = 100°C), a small refractive index n, and a large absorption coefficient α in the visible. The absorption thresholds as represented by the optical gap $E_g$ are in the range of 1,58 eV and 2,2 eV. These results are close to those already published in the literature. For low $T_s$, hydrogen atoms saturate most of dangling bonds existing in the deposited layers. This is shown through the correct values obtained for the optical gap $E_g$ and the Urbach energy $E_u$. For higher $T_s$, hydrogen begins to out-diffuse and the optical performances of the layers start decreasing. After thermal annealings at moderate temperatures such as 340°C or 400°C, a high quantity of hydrogen incorporated in the layers out-diffuses. After thermal annealing at 570°C, hydrogen out-diffusion is higher but some partial re-crystallisation of the layers should happen. This is why a clear improvement in their optical parameters has been observed.

## ACKNOWLEDGEMENTS

This paper has been prepared on the basis of experimental results obtained essentially by the co-author B. M. Aka. It has been structured, enriched by literature data and written during the scientific stay of the first author A. Barhdadi, as Senior Associate Professor, at the Abdus Salam International Centre for Theoretical Physics (ICTP). This author would like to thank the Director and staff of the Centre for their kind hospitality, efficient assistance and great support. He also wishes to thank a lot Professor G. Furlan, Head of ICTP-TRIL Programme, for his precious scientific cooperation and paper referring. Special thanks are also addressed to Profs. D. Sayeh and M. Abd-Lefdil, respectively the former and the present Directors of ″Physique des Matériaux″ Laboratory, Faculty of Sciences, University Mohammed V-Agdal, Rabat, for their great scientific and technical contributions.